\documentclass[journal=jacsat,manuscript=article]{achemso}

\newcommand{\be}{\begin{equation}}
\newcommand{\ee}{\end{equation}}
\newcommand{\bea}{\begin{eqnarray}}
\newcommand{\eea}{\end{eqnarray}}
\def\lb{\label}

\usepackage[version=3]{mhchem} % Formula subscripts using \ce{}

\author{I. Maccari}
\affiliation{Institute for Theoretical Physics, ETH Zurich, 8093 Zurich, Switzerland}
\email{imaccari@phys.ethz.ch }
\author{L. Benfatto}
\affiliation{Department of Physics, Sapienza University of Rome, P.le A. Moro 2, 00185 Rome, Italy}
\alsoaffiliation{Institute for Complex Systems (ISC-CNR), UOS Sapienza, P.le A. Moro 5, 00185 Rome, Italy}
\author{C. Castellani}
\affiliation{Department of Physics, Sapienza University of Rome, P.le A. Moro 2, 00185 Rome, Italy}
\alsoaffiliation{Institute for Complex Systems (ISC-CNR), UOS Sapienza, P.le A. Moro 5, 00185 Rome, Italy}

\author{J.Lorenzana}
\affiliation{Institute for Complex Systems (ISC-CNR), UOS Sapienza, P.le A. Moro 5, 00185 Rome, Italy}

\author{C. De Michele}
\affiliation{Department of Physics, Sapienza University of Rome, P.le A. Moro 2, 00185 Rome, Italy}
\email{cristiano.demichele@uniroma1.it}

\title{Fragile-to-strong glass transition in two-dimensional vortex liquids}

\abbreviations{IR,NMR,UV}

\begin{document}

% \begin{tocentry}

% Some journals require a graphical entry for the Table of Contents.
% This should be laid out ``print ready'' so that the sizing of the

% Inside the \texttt{tocentry} environment, the font used is Helvetica
% 8\,pt, as required by \emph{Journal of the American Chemical

% The surrounding frame is 9\,cm by 3.5\,cm, which is the maximum
% permitted for  \emph{Journal of the American Chemical Society}
% graphical table of content entries. The box will not resize if the
% content is too big: instead it will overflow the edge of the box.

% This box and the associated title will always be printed on a
% separate page at the end of the document.

% \end{tocentry}

\begin{abstract}

The fragile-to-strong glass transition is a fascinating phenomenon that still presents many theoretical and experimental challenges. A major one is how to tune the fragility of a glass-forming liquid. 
Here, we study a two-dimensional (2D) system composed of vortices in a superconducting film, which effectively behaves as a 2D glass-forming liquid.
We show that the kinetic fragility in this system can be experimentally varied by tuning a single parameter: the external magnetic field $H$ applied transversely to the film.
This conclusion is supported by the direct comparison between the analysis of experimental measurements in an amorphous MoGe superconducting film and Monte Carlo simulations in a disordered XY model, that captures the universality class of the two-step melting transition.  We show that by increasing disorder strength a fragile-to-strong transition is induced, in close similarity with the experimental findings in a magnetic field. Our numerical results shed light on the evolution of the dynamical heterogeneity from a fragile to strong glass, as due to the subtle interplay between caging effects arising from hexatic order and strong random pinning.

\end{abstract}

\section{Introduction}

The slowing down of the dynamics of a glass-forming liquid as it approaches the glass transition is a fascinating phenomenon still poorly understood. Kauzmann~\cite{Kauzmann1948} was the first to realize that if equilibrium could be maintained during cooling, at some temperature $T_K$ the entropy of the liquid becomes as low as that of an ordered state. However, the dynamics become so slow before $T_K$ can be attained that equilibration is no longer possible and the system becomes a glass at a temperature $T_g > T_K$.  How the dynamics slow down differs from system to system\cite{Angell1995}. While some liquids termed “strong” display an  Arrhenius behaviour,
% {\color{green}--- 
where the viscosity $\eta(T)$ is proportional to $\exp(B/T)$,
%---} 
which can be easily interpreted in terms of energy barriers, others termed “fragile” show a much more dramatic increase in $\eta(T)$, which is well described by the Vogel-Fulcher-Tamman (VFT) law~\cite{Vogel1921, Fulcher1925, Tamman1926}, i.e. $\eta(T) \sim \exp(B/(T-T_0))$. Despite being a phenomenological law, the temperature $T_0$ at which the viscosity diverges indicates the presence of a dynamical phase transition which is expected to coincide with the thermodynamic transition occurring at $T_K$~\cite{cavagnaSupercooledLiquidsPedestrians2009,Ruocco2004}. 
This extraordinary increase in viscosity has often been attributed to dynamical heterogeneity which can emerge 
as a result of particle-particle interactions.
In colloids, the observed fragility variation~\cite{raymondp.seekellRelationshipParticleElasticity2015a,nigroDynamicalBehaviorMicrogels2017, nigroRelaxationDynamicsSoftness2020} has been microscopically related to the particle potential softness~\cite{mattssonSoftColloidsMake2009, vanderScheerFragility2017, Yu2022} and particle deformation~\cite{gnanMicroscopicRoleDeformation2019}. 
In two-dimensional (2D) systems, dynamical heterogeneity can also originate from the two-step melting transition from the solid to the liquid phase, as predicted by the Berezinskii-Kosterlitz-Thouless-Halperin-Nelson-Young (BKTHNY) theory \cite{berezinskyDestructionLongrangeOrder1972, kosterlitzOrderingMetastabilityPhase1973, kosterlitzCriticalPropertiesTwodimensional1974, halperinTheoryTwoDimensionalMelting1978, nelsonDislocationmediatedMeltingTwo1979, youngMeltingVectorCoulomb1979}.  This is indeed associated with increased hexatic correlations in the intermediate liquid phase, which lacks long-range translational order but retains orientational correlations~\cite{zangiCooperativeDynamicsTwo2004,shibaStructuralDynamicalHeterogeneities2009, kawasakiStructuralSignatureSlow2011,Krauth2011,Krauth2015, meerDynamicalHeterogeneitiesDefects2015, tahGlassTransitionSupercooled2018}.

\begin{figure}[b!]
    \centering    \includegraphics[width=\linewidth]{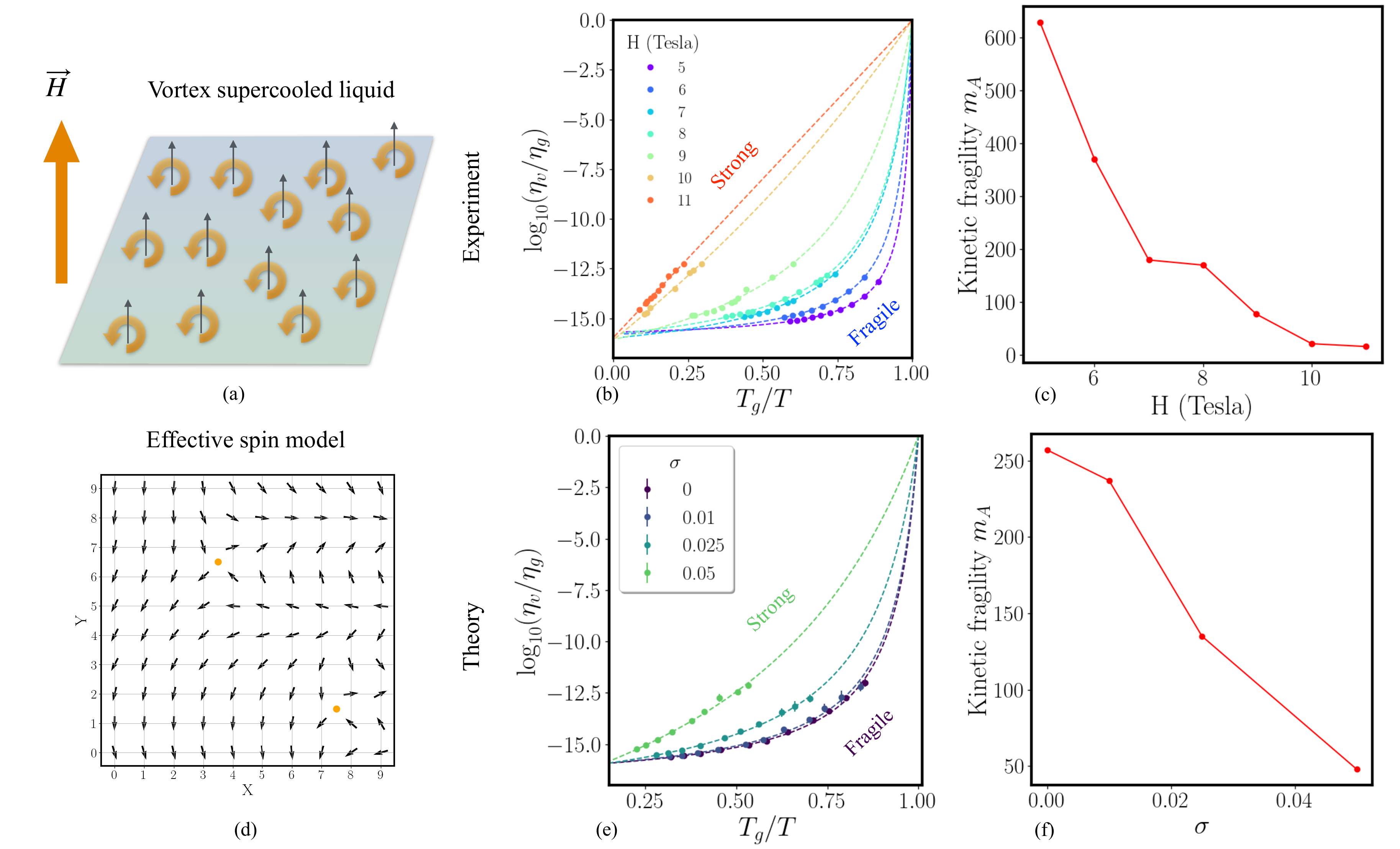}
    \caption{(a) Schematic representation of vortex supercooled liquid forming in a SC thin film. The black arrows indicate the magnetic flux quanta entered into the system, leading to the formation of vortices in the gauge-invariant SC phase, indicated as orange circular arrows. (b) Angell's plot for the viscosity of the 2D vortex lattice viscosity forming in an amorphous MoGe thin film. The vortex viscosity is obtained from the low-temperature resistivity, $\rho(T) = \Phi_0 H/\eta_v(T)$, for different values of the applied transverse magnetic field $H$ (T). The experimental data for the resistivity are those presented in~\cite{maccariTransportSignaturesFragile2023}. As one can see, by increasing the applied field the system transitions from a fragile to a strong glass. 
    (c) The extracted dynamic fragility of the glass-former vortex liquid.
    (d)Snapshot of a portion of the spin grid numerically simulated.  The vortices emerge as topological excitations of the superconducting phase and are shown here as orange dots. 
    (e) Angell’s plot for the viscosity obtained via Monte Carlo simulations on a 2D vortex lattice for different disorder strengths $\sigma$ and fixed magnetic field ${H}$. By increasing $\sigma$ the system transitions from a fragile to a strong glass behaviour. 
    (f) Kinetic fragility of the simulated 2D vortex liquid as a function of the quenched disorder strength $\sigma$.
    For both panels (b) and (e),  the dashed lines correspond to a VFT fit of the data. 
    }
    \label{Angell_exp}
\end{figure}

Along with colloids, type-II superconducting (SC) films offer a promising platform to study the fragility of two-dimensional glass-forming liquids and shed light on its microscopic origin. 
When a magnetic field ${H}$ is applied perpendicularly to the SC film, and it exceeds a critical threshold called $H_{c1}$, it starts to penetrate in the form of quantized superconducting vortices which behave as interacting classical particles, see Fig. \ref{Angell_exp}(a). 
Each vortex carries a quantum flux $\Phi_0 = h/2e$, where $h$ is Planck's constant and $e$ is the electric charge, the magnitude of the applied field thus determines the vortex density induced in the system, with $n_v = H/\Phi_0$. 
These vortices effectively behave as 2D Coulomb charges whose free energy landscape is ultimately determined by the interplay between their interaction potential, whose coupling constant is set by the superfluid density $n_s$, and the quenched disorder, which may arise from atomic-scale point defects and acts as a pinning force.
At low temperatures and weak disorder, a vortex solid forms as a 2D Bragg glass~\cite{giamarchiElasticTheoryFlux1995} and the system is superconductive. The superconducting (SC) transition to a metallic phase coincides with the vortex solid melting into a vortex liquid.
In a recent work~\cite{maccariTransportSignaturesFragile2023} , this framework was applied to study the melting of a weakly-pinned 2D vortex lattice in amorphous MoGe superconducting thin films.
In this system, the two-step melting transition via a hexatic liquid phase was successfully characterized experimentally by combining transport and scanning tunnelling spectroscopy measurements~\cite{royMeltingVortexLattice2019, duttaExtremeSensitivityVortex2019}.
The subsequent theoretical analysis of transport data in an extended region of $H$ and $T$ has introduced a novel paradigm for interpreting magnetotransport measurements.
That analysis relies on the fact that the linear resistivity is directly related to the dynamics of vortices being $ \rho(T) = (h/2e)^2 n_v/\eta_v(T)$, with $\eta_v(T)$ the vortex viscosity. 
In Ref.~\cite{maccariTransportSignaturesFragile2023} , the linear resistivity $\rho(T)$ was directly fitted with a VFT law and $T_0$ was extracted as a function of $H$.
That allowed the identification of
fragile-glass signatures in the hexatic phase and a crossover from fragile to strong glass behaviour by increasing  $H$, signalled by the vanishing of $T_0$ and the restoration of an Arrhenius behaviour. 
Nevertheless, 
a theoretical model explaining the observed experimental behaviour was missing, and many of the questions that this experiment raised, related to the role played by disorder, magnetic field and vortex interactions, remained unanswered.

Here, we provide a new theoretical analysis of the experimental data. We present Angell's plot of the experimental vortex viscosity and extract the fragility of the vortex liquid as a function of $H$, see Fig.\ref{Angell_exp}(b)-(c).
This result demonstrates how 2D vortex liquids forming in thin SC films offer an unprecedented opportunity to study the fragile-to-strong glass transition by varying a single parameter within the same physical system.
At the same time, we propose a theoretical model able to explain the fragile-to-strong glass transition experimentally observed in terms of an effective increase of quenched disorder with $H$.

Monte Carlo numerical simulations on a 2D XY model in the presence of a transverse magnetic field show how by varying the level of quenched disorder, parametrized by $\sigma$, the vortex liquid fragility decreases similar to what experimentally observed by varying $H$. The numerical results are summarized in Fig.\ref{Angell_exp} (e)-(f)
Our work presents a novel analysis of the experimental phase diagram discussed in~\cite{maccariTransportSignaturesFragile2023} while addressing the more general question about the fate of the fragile-glass signatures of the hexatic phase by increasing the disorder strength. 

This work aims at building a bridge between the soft matter community and the community working on superconductivity and strongly correlated systems.

\section{Results and discussion}
\textbf{Experimental data.}

Starting from the linear resistivity measured at different magnetic fields in an amorphous MoGe thin SC film~\cite{maccariTransportSignaturesFragile2023}, we extract the vortex viscosity as $\eta_v(T) = (h/2e)^2 n_v/ \rho(T)$ and plot it in Fig.\ref{Angell_exp} (b) according to Angell's plot perscription~\cite{angellRelaxationLiquidsPolymers1991}.
{Conventional glass-forming liquids, at temperatures far from glass transition,
typically have a viscosity $\eta_\infty$ of the order of $10^{-2} \mathrm{Pa} \cdot \mathrm{s}$. At the glass critical temperature,
when their relaxation time is of the order of $100$ s, their viscosity reaches values
of the order of $10^{14} \mathrm{Pa} \cdot \mathrm{s}$. This means that the viscosity changes by $16$ orders of magnitude from the high-temperature regime to the glass critical point. According to this convention, here we define the glass transition temperature, $T_g$, as the temperature at which the viscosity becomes $16$ orders of magnitude larger than the extrapolated viscosity at large temperatures, i.e. $\eta_g= 10^{16} \cdot \eta_v(T \to \infty) = 10^7 \mathrm{Pa} \cdot \mathrm{s}$.
In Fig.\ref{Angell_exp} (b), different values of the applied magnetic field are shown to demonstrate the transition from a fragile to a strong vortex glass by increasing the magnetic field $H$.
Note that in ref~\cite{maccariTransportSignaturesFragile2023}, the experimental resolution for the resistivity was $\rho_{min}=3.6 \times 10^{-4} \mathrm{m \Omega} \cdot \mathrm{cm}$ 
%$R_{min} \sim 0.5 \Omega$
corresponding to $\eta_{max} \sim 10^{-6} \mathrm{Pa} \cdot \mathrm{s}$. Although experimentally the regime of viscosity close to the conventional $\eta_g$ is not accessible (i.e.  $\eta_{max} \ll \eta_g$), the difference between fragile and strong behaviour clearly emerges from the data.

From the experimental values of the vortex viscosity, we then extract the value of the vortex kinetic fragility which measures the \emph{steepness} of the temperature dependence of the liquid viscosity at the glass transition $T_g$~\cite{plazekCorrelationPolymerSegmental1991, bohmerCorrelationsNonexponentialityState1992}

\begin{equation}
m_A= \frac{d[\log(\eta_v(T)/ \eta_{g})]}{d[T_g/T]} {\Large|}_{T=T_g}.
\label{fragility}
\end{equation}

The resulting kinetic fragility, $m_A$, as a function of the magnetic field is shown in Fig.\ref{Angell_exp} (c).
In real glass-former systems, the kinetic fragility varies over one order of magnitude ranging from 17 in strong glasses like silica to 200 in fragile glasses~\cite{angellRelaxationLiquidsPolymers1991}. A comparable variation is found for the kinetic fragility of a vortex-supercooled liquid by varying a single external knob, $H$.

Note that our conclusions do not depend on the value of   $\eta_g$ used to define the glass. Indeed, using  $\eta_g=10^{-6} \mathrm{Pa} \cdot \mathrm{s}$ the Angell plot shows qualitatively the same transition from fragile to strong. On the other
hand, absolute values of the kinetic fragility depend on the choice of
$\eta_g$. For example, taking $\eta_g=10^{-6} \mathrm{Pa} \cdot \mathrm{s}$   the kinetic fragility becomes more than an order of magnitude smaller but follows the same trend.

 \textbf{The theoretical model.} To understand the origin of the fragile-to-strong glass transition observed in 2D vortex liquids, we perform Monte Carlo (MC) numerical simulations on an effective spin model for 2D superconductors. 
To avoid spurious effects due to the incommensurability between the vortex lattice and the numerical square grid\cite{maccariTransportSignaturesFragile2023}, see Fig.\ref{Angell_exp} (d), we keep the total number of vortices constant, while varying the level of quenched disorder, which effectively increases by increasing the magnetic field $H$.
Indeed, as $H$ increases in the experiments, the energy required to break Cooper pairs decreases, leading to a reduction in the superconducting density $n_s$. Consequently, the energy scale associated with vortex-vortex interactions (proportional to $n_s$) decreases compared with the energy scale of random pinning, which is directly coupled to the vortex density~\cite{giamarchiElasticTheoryFlux1995}. 
That is why in thin SC films, the increase of the vortex density with $H$ does not have the same effects of increasing the particle density in colloidal systems~\cite{saika-voivodFragiletostrongTransitionPolyamorphism2001}. In the following, we will show how it is precisely the effective increase in quenched disorder which drives the fragile-to-strong glass transition in 2D vortex liquids.

To access both the superconducting response of the system and the vortex lattice dynamics, we study an effective spin model for the SC phase, where vortices appear as topological phase excitations, see Fig.\ref{Angell_exp} (d).
The 2D $XY$ model, whose Hamiltonian on a discrete lattice reads  

\be
H_{\mathrm{XY}}=- J \sum_{i,\mu=\hat{x}, \hat{y}} \cos(\theta_i-\theta_{i+\mu}),
\label{xymodel}
\ee
describes the Josephson-like interaction between the nearest-neighbour SC islands with phase $\theta_i$ and fixed SC density $n_s \propto J$.

At zero external magnetic field, the superconducting transition belongs to the Berezinskii-Kosterlitz-Thouless (BKT) universality class~\cite{berezinskyDestructionLongrangeOrder1972, kosterlitzOrderingMetastabilityPhase1973, kosterlitzCriticalPropertiesTwodimensional1974}, where vortices and antivortices are thermally nucleated and their unbinding at the critical point destroys the superconducting order.
 
Here, we study a 2D $XY$ model in the presence of a transverse magnetic field and quenched disorder so that Eq.\eqref{xymodel} becomes
% The Hamiltonian on a discrete lattice reads:
%
\be
H_{\mathrm{XY}}=- \sum_{i,\mu=\hat{x}, \hat{y}} J^{\mu}_i \cos(\theta_i-\theta_{i+\mu}+F^\mu_i).
\label{xymodel_diso}
\ee
The SC phase of the condensate on the site $i$, $\theta_i$, is minimally coupled to the external magnetic field via the Peierls substitution %\CDMnote{citazione?} è molto fondamentale
 $F_i^\mu=\frac{2\pi}{\Phi_0}\int_{r_i}^{r_{i+\mu}}A^\mu_i\cdot dr_{\mu}$. The presence of a finite transverse magnetic field, $H\hat z=\vec{\nabla}\times \vec{A}$, frustrates the ferromagnetic coupling between neighbouring sites inducing in the system a finite number of vortices with a vorticity defined by the sign of $H\hat{z}$. Each vortex carries a quantum of flux $\Phi_0$, so that the total number of vortices in the ground state is $N_v= f L^2$, where $L$ is the linear size of the system and the filling fraction $f= Ha^2/\Phi_0$ is given by the magnetic flux passing through a unitary plaquette of linear size $a=1$.  
Similarly to a previous work~\cite{maccariUniformlyFrustratedXY2021},  the presence of a quenched disorder is embedded in the Josephson couplings $J_i^{\mu}$ between nearest neighbouring sites $i$ and $i+\hat{\mu}$, with $\hat{\mu}= \hat{x}, \hat{y}$. For each link $i, i+\hat{\mu}$, $J_i^{\mu}$ is extracted randomly from a Gaussian distribution with mean $\bar{J}=1$ and variance $\sigma$. The disorder strength is controlled by $\sigma$.
Differently from the $f=0$ case, the phase transition is now driven by the melting of the 2D vortex lattice whose nature strongly depends on the filling fraction $f$ and quenched disorder. 

The Monte Carlo (MC) simulations of the model \eqref{xymodel_diso} are carried out on a square grid of lattice spacing $a=1$, linear size $L=56$ and periodic boundary conditions. We fix the external magnetic field intensity to $f= 1/L$, resulting in $N_\mathrm{v}=fL^2=56$ vortices. 
For each temperature and disorder level, we compute the mean value and the statistical error of a given observable by averaging both on the MC steps and on $N_{samples}$ independent realizations of disorder.
More details on the MC simulations are reported in the Methods section.

As already mentioned, by treating vortices as topological excitations of the phase field, the model \eqref{xymodel_diso} allows us to study both the vortex lattice ordering and the superfluid response of the system. 
Furthermore, in this work, we characterize the static and dynamic properties of the vortex lattice for each level of disorder studied. This allows us to assess the fragility of the lattice as a function of the quenched disorder and gain insights on the experimental phase diagram discussed in~\cite{maccariTransportSignaturesFragile2023}.

\textbf{Static properties}. To assess the superconducting phase transition we compute the superfluid stiffness $J_s^{\mu}$, defined as the system response to a uniform twist of the gauge-invariant phase, $(\theta_i-\theta_{i+\mu}+F^\mu_i) \to (\theta_i-\theta_{i+\mu}+F^\mu_i) +\Delta_{\mu}$, along a given direction $\mu$

\begin{equation}
    J_s^{\mu}= -\frac{1}{L^2} \frac{\partial^2 \ln Z(\Delta_{\mu})}{\partial \Delta_{\mu}^2} \Big|_{\Delta_{\mu}=0},
\end{equation}
where $Z(\Delta_{\mu})$ is the partition function of the model \eqref{xymodel_diso}.
We compute the superfluid stiffness (see Supporting Information for its explicit expression) along both $\hat{x}$ and $\hat{y}$ and we label $J_s = {(\langle J_s^x \rangle  + \langle J_s^y\rangle )}/{2}$. Here and in what follows, $\langle \dots \rangle$ stays both for the thermal average over the MC steps and the $N_{samples}=15$ independent disorder configurations (see Methods for more details). 
 
\begin{figure}[h!]
\centering
\includegraphics[width=
\linewidth]{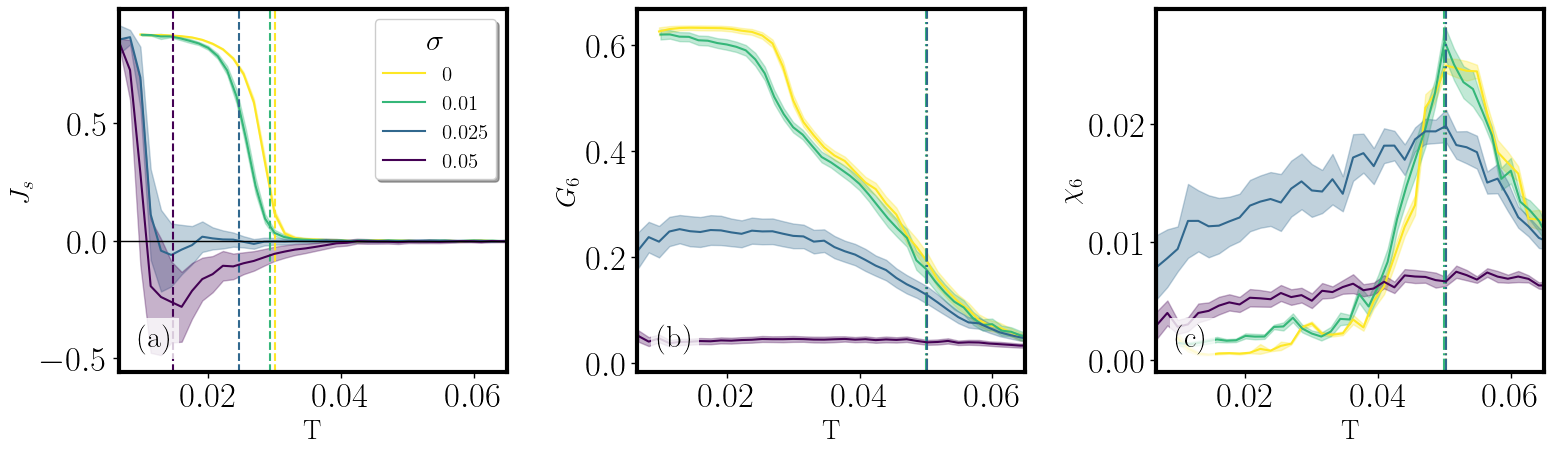}
    \caption{Static properties of the model \eqref{xymodel_diso} with linear size $L=56$ and uniform filling fraction $f=1/L$ for different disorder strengths, as encoded in the standard deviation $\sigma$ for the Gaussian distribution of the couplings. (a) Temperature dependence of the superfluid stiffness $J_s$. The vertical dashed lines indicate the values of $T_0$ extracted from the VFT fit of the vortex diffusion coefficient at each disorder level. (b)-(c) Temperature dependence of the orientational order parameter $G_6$ and its susceptibility $\chi_6$. The vertical dot-dashed lines indicate $T_{hex}\simeq0.05$, extracted for $\sigma<0.05$ from the peak of the hexatic susceptibility.  The error bars are indicated as shaded regions around the lines. }
    \label{fig:Js_G6}
\end{figure}

The temperature dependence of $J_s$ is shown in Fig.\ref{fig:Js_G6} (a) for different levels of disorder, including the clean case where $\sigma=0$. By increasing the disorder strength, the zero temperature value of the superfluid stiffness  $J_s(T=0)$ stays almost unchanged, while the critical temperature at which $J_s\neq 0$ strongly decreases with $\sigma$. At the same time, by increasing $\sigma$ the thermalization of the system becomes more challenging as reflected by the large error bars and the slightly negative values of $J_s$ approaching the critical temperature $T_c$.

% \twocolumngrid\
%
At the same time, the vortex pinning induced by the quenched disorder competes with the vortex-vortex interaction affecting the vortex lattice order. To investigate the impact of quenched disorder on the hexatic order, we compute the six-fold orientational order parameter $G_6$.
To this aim, we first determine the position of the vortices from the gauge-invariant phase circulation around each unitary plaquette of the square spin grid (see the Supporting Information for more details), and we identify the nearest neighbours of each vortex via a Delaunay triangulation of the vortex lattice. Finally, we compute the local orientational order $\psi_{6j}$ relative to the $j-$th vortex as
\be
\psi_{6j}= \frac{1}{N_j} \sum_{k=1}^{N_j} e^{6 i \theta_{jk}}, 
\label{psij}
\ee
where $N_j$ is the number of its nearest neighbours, and $\theta_{jk}$ is the angle that the bond connecting the two neighbouring vortices $j$ and $k$ forms with respect to a fixed direction in the plane.

The orientational order parameter $G_6$ is then obtained from $\Psi_6= \frac{1}{N_\mathrm{v}} \sum_{j=1}^{N_\mathrm{v}} \psi_{6j} $
\be
\lb{g6}
G_6=\langle \Psi_6 \rangle, 
\ee
and its corresponding susceptibility as
\be
\lb{chi6}
\chi_6=\langle \Psi_6^2 \rangle - \langle \Psi_6 \rangle^2 . 
\ee

The temperature dependence of $G_6$ and $\chi_6$ for different disorder strengths are shown in Fig.\ref{fig:Js_G6} (b)-(c).  
At low disorder levels,  the temperature where $G_6$ becomes finite (i.e. where the system undergoes a hexatic-to-isotropic liquid transition) weakly depends on $\sigma$, in agreement with previous results on 2D colloidal systems with random pinning~\cite{deutschlanderTwoDimensionalMeltingQuenched2013, Gaiduk2019, Shankaraiah2019}. As long as the hexatic phase exists, indeed, the hexatic critical temperature, extracted from the peak in $\chi_6$, remains that of the clean system being $T_{hex} \simeq0.05$, see the dot-dashed lines in Fig.\ref{fig:Js_G6} (b)-(c).  On the other hand, the zero-temperature value of the hexatic order parameter, i.e. $G_6(T=0)$, strongly depends on $\sigma$, and it becomes vanishingly small at $\sigma=0.05$, where also the hexatic susceptibility $\chi_6(T)$ does not show a clear peak. 
The vanishing of $G_6(T=0)$ for a large enough value of $\sigma$ signals the destruction of the hexatic phase and the appearance of a disordered SC vortex solid.

\textbf{Dynamical properties.} To investigate further how the 2D melting transition evolves by increasing the disorder strength, we look at the dynamic properties of the system, focusing on its slowing down as it approaches the superconducting phase from the high-temperature liquid phase. 
We tracked the position of each vortex (see the Supporting Information for more details)
in time and space and extracted several dynamical observables for different temperatures and disorder levels.

The results for the vortex lattice dynamical autocorrelation function, and the non-gaussian parameter, which measures the degree of dynamic heterogeneity, are discussed and shown in Fig.1-2 in the Supporting Information.
Here, we focus on the mean-square displacement $\langle \Delta r^2(t) \rangle$
and the vortex viscosity $\eta_v $ from which we can directly characterize the vortex lattice fragility as a function of the quenched disorder. 
As before, $\langle \dots \rangle$ indicates the average both over the MC steps and $N_{samples}=10$ independent realization of disorder.

The resulting curves of $\langle \Delta r^2(t) \rangle$ for different temperatures and disorder levels are shown in Fig.~\ref{fig:msd_eta} (a)-(d).
At large temperatures, the system shows just two time-scale regimes: a sub-diffusive regime at short times, for distance comparable with the lattice spacing of
%to the presence of a 
the discrete numerical grid, and the expected diffusive regime at larger times, where $\langle \Delta r^2(t) \rangle \sim D_v t$. 
 However, for temperatures lower than $T_{hex}$, indicated in Fig.\ref{fig:msd_eta} (a)-(d) with a grey colour, a second subdiffusive regime appears, signalling the emergence of a heterogeneous dynamic.
  At low disorder, where a hexatic phase exists, the heterogeneity of the vortex dynamics can be understood in terms of the quasi-long-range orientational order between vortices that prevents them from moving isotropically in the system. 
  
\begin{figure}[b!]
    \includegraphics[width=\linewidth]{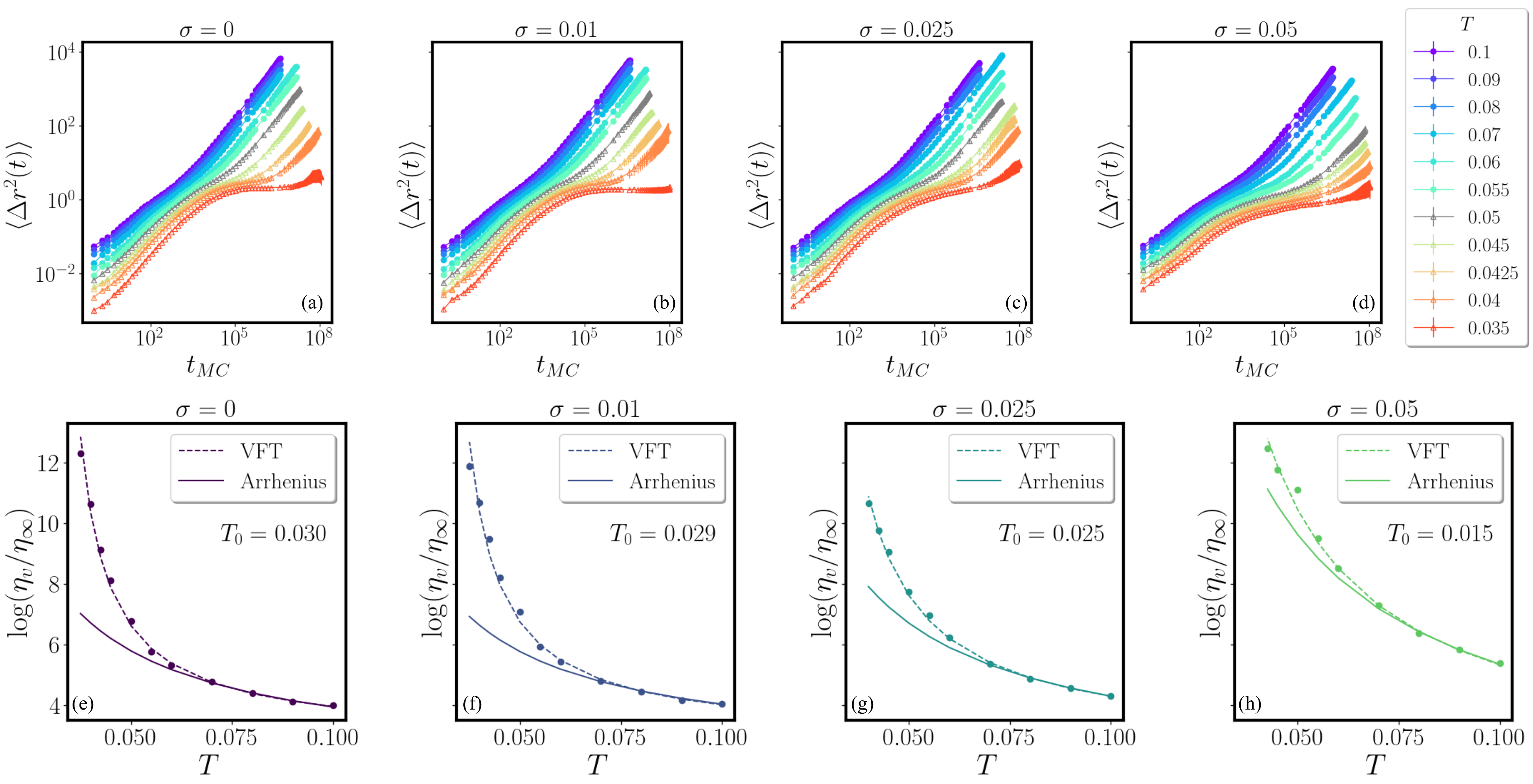}
    \caption{Upper panels - Vortex mean-square displacement $\langle \Delta r^2(t) \rangle$ at different temperatures and for three different disorder levels: (a) $\sigma=0$, i.e. clean case; (b) $\sigma=0.01$; (c) $\sigma= 0.025$; and (d) $\sigma = 0.05$. 
    For $\sigma<0.05$, the grey-coloured temperature indicates the critical temperature at which the hexatic liquid phase appears. For $\sigma=0.05$, it signals the appearance of a second subdiffusive regime indicative of emerging heterogeneous dynamics despite the lack of a marked hexatic ordering as signalled by the absence of a peak in $\chi_6(T)$ for this disorder level, see Fig.\ref{fig:Js_G6} (c).  Lower panels - Temperature dependence of the natural logarithm of the vortex viscosity, $\eta_v= k_\mathrm{B}T/D_v$, renormalized by $\eta_\infty= \eta_v(T\to \infty)$, for three different disorder levels: (e) $\sigma=0$, i.e. clean case; (f) $\sigma=0.01$; (g) $\sigma= 0.025$; and (h) $\sigma = 0.05$. The dashed lines in the panel (e)-(h) indicate the fit of $\eta_v(T)$ obtained using the VFT functional form, i.e. $\eta_v(T) = A \exp(-\frac{B}{T-T_0})$, while the continuous lines indicate the Arrhenius fit, i.e. $\eta_v(T)= \tilde{A} \exp(-\frac{\tilde{B}}{T}) $.}
    \label{fig:msd_eta}
\end{figure}

That is also the case for $\sigma=0.05$, see Fig.\ref{fig:msd_eta} (d), where a second subdiffusive regime also emerges around $T\simeq0.05$ despite the 
%lack of the hexatic phase
presence of a vanishingly small orientational order parameter, see Fig.~\ref{fig:Js_G6} (c). 
The presence of heterogeneous dynamics at this disorder strength is also confirmed by the non-Gaussian parameter $\alpha_2(t)$ (see Fig.~\ref{fig:alpha}). 
This should be understood as a crossover regime, where a \emph{pinning-induced} cage takes the place of the hexatic cage forming at lower disorder levels. Eventually, this will result in a transition to a strong glass behaviour at increasing disorder, as will be evident from the following analysis of the vortex viscosity.

From the large-time asymptotic behaviour of $\langle \Delta r^2(t) \rangle$ we extract the vortex diffusion coefficient
as 
%\be
$D_v = \frac{1}{4} \lim_{t \to \infty} { \langle \Delta {r}^2(t) \rangle}/{t}$
%\ee
and compute the vortex viscosity via $\eta_v = \frac{k_\mathrm{B}T}{D_v a}$. Notice that the time is in units of Monte Carlo steps and, as customary, we assume that the {
long-time (glassy) fictitious MC dynamics coincide with the physical dynamics upon a suitable scaling factor~\cite{Kob2007}. Because of this, dynamical quantities such as the diffusion constant or the vortex viscosity are meaningful only as relative values.

The temperature dependence of $\eta_v$, renormalized by $\eta_g$, are shown for different disorder levels in Figures \ref{fig:msd_eta} (e)-(h). 

At low disorder, similarly to the clean case, see Fig.\ref{fig:msd_eta} (e), the vortex viscosity significantly deviates from the Arrhenius behaviour at low temperature following instead the phenomenological VFT law 

\be
\eta_v = \eta_{\infty} \exp(- \frac{B}{T-T_0}).
\label{etav_VFT}
\ee
where $\eta_\infty=\eta(T\to \infty)$.

As the level of disorder increases, the deviation becomes less pronounced, resulting in smaller values of $T_0$ which eventually vanishes at even larger $\sigma$.

Note that being $T_0$ the temperature at which the vortex viscosity diverges, one would expect it to coincide with the temperature, $T_{c,J_s}$ at which  $J_s\neq 0$.
At low disorder, $\sigma < 0.025$, we find a good agreement between $T_{c,J_s}$ and $T_0$,  see Fig.\ref{fig:Js_G6} where $T_0$ are indicated as vertical dashed lines. At larger disorder, due to a significant increase of the relaxation time as reflected by larger error bars in the computed $J_s(T)$, it becomes harder to assess the coincidence between $T_0$ and $T_{c,J_s}$.

The vanishing of $T_0$ as a function of the quenched disorder, see Fig.\ref{fig:TgT0_etav}, indicates a transition from a fragile to a strong glass behaviour which appears even more evident by looking at the corresponding Angell's plot of $\log_{10}(\eta_v/\eta_{g})$ vs $T_g/T$ shown in Fig.\ref{Angell_exp} (e).
Here, $T_g$ is defined as the temperature at which the viscosity reaches the threshold of $\eta_g= \eta_\infty \cdot 10^{16}$, with $\eta_\infty \sim 10^{-2}$ from our MC simulations. 
The glass temperature $T_g$ is plotted alongside $T_0$ as a function of the disorder level in Fig.\ref{fig:TgT0_etav}.

\begin{figure}[h!]
\includegraphics[width=0.5\linewidth]{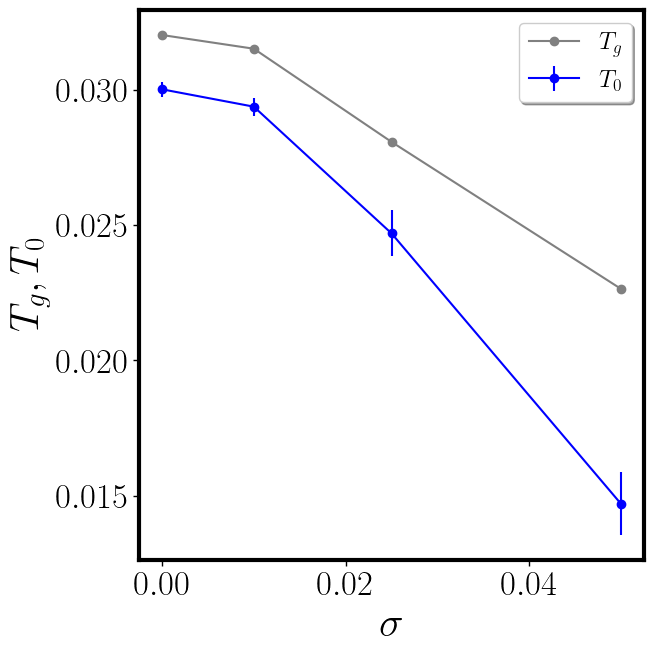}
    \caption{(a) Temperature $T_0$, obtained from the VFT fit in Eq.\eqref{etav_VFT}, as a function of the disorder strength $\sigma$ plotted together with the glass critical temperature $T_g$ defined as before as $\eta_v(T_g)= \eta_\infty \cdot 10^{16}$. }
\label{fig:TgT0_etav}
\end{figure}

As already discussed, a more quantitative measure of the deviation from Arrhenius's behaviour is provided by the kinetic fragility $m_A$, defined in Eq.\eqref{fragility},
%
% \begin{equation}
% m_A= \frac{d[\log(\eta_v(T)/ \eta_{g})]}{d[T_g/T]} {\Large|}_{T=T_g},
% \label{fragility}
% \end{equation}
%
which is shown in Fig.~\ref{Angell_exp}(f).
Both Angell's plot and the kinetic fragility obtained from our MC simulations are in qualitative agreement with the experimental Angell's plot and kinetic fragility shown in Fig.\ref{Angell_exp} (b)-(c) respectively. 

Hence, according to our numerical findings, the quenched disorder plays a role analogous to that of the magnetic field in the experiments, thus clarifying the nature of the fragile-to-strong glass transition experimentally observed in~\cite{maccariTransportSignaturesFragile2023}.

In the experiments, at low magnetic field $H$, the effect of the quenched disorder is small and a hexatic vortex liquid appears. That is characterized by a collective motion of vortices, needed to preserve orientational correlations, which results in a high kinetic fragility and a super-Arrhenius behaviour. 
The effective increase of quenched disorder strength with $H$ destroys the hexatic liquid phase and the vortex motion loses its collective character as the role of random pinning becomes dominant. Eventually, the Arrhenius behaviour is recovered and the vortex liquid becomes a strong glass.

\section{Conclusions}

In this work, we have provided a theoretical framework to explain the fragile-to-strong glass transition occurring in an amorphous superconducting film by increasing the transverse magnetic field $H$. Starting from the resistivity data presented in~\cite{maccariTransportSignaturesFragile2023}, we extracted the vortex liquid viscosity and its kinetic fragility. 
The analysis of the experimental data, as shown in Fig.\ref{Angell_exp} (b), demonstrates that tuning a single external parameter, $H$, varies the vortex liquid fragility. 

To gain a deeper insight into the fragile-to-strong glass transition, we studied an effective spin model for the SC phase via Monte Carlo simulations.
Since in SC thin films increasing $H$ corresponds to increasing the effective quenched disorder, we studied the transition at fixed $H$ by increasing the disorder strength.
We showed how the level of the effective quenched disorder can determine the nature of the glass-forming liquid. 
Indeed, alongside a progressive destruction of the orientational order, the increase of the disorder makes the glass stronger. 
That can be understood in terms of disorder potential energy barriers competing with the vortex-vortex interactions. At low disorder, the glass-former vortex liquid exhibits a hexatic liquid phase resulting in a fragile glass behaviour. By increasing the disorder, the vortex liquid orientational order gets progressively disrupted by random pinning until the vortex dynamic becomes the dynamic of individual vortices controlled by a single time scale associated with the disorder potential, so that the vortex supercooled liquid recovers an Arrhenius-like activated dynamics. Interestingly, in the high-disorder regime, we observe a logarithmic behaviour in the long-time decay of the intermediate scattering function (see Fig.1 of the Supporting Information), which reflects the competition between two distinct mechanisms slowing down the vortex relaxation, i.e. the (hexatic) caging and pinning~\cite{Sentjabrskaja2016AnomalousDO}.

In the present work, we used a weak Gaussian distributed quenched disorder. 
However, other kinds of disorder may affect the vortex dynamics in different ways. Exploring how different quenched disorders modify the observed fragile-to-strong glass transition is an interesting question that we will face in a future study.

A related question is the effect of different types of quenched disorder on the configurational entropy $S_c$. 
Indeed, quenched disorder 
may act by blocking out some particles, reducing the number of possible states and thus decreasing $S_c$~\cite{cammarotaIdealGlassTransitions2012, williamsExperimentalDeterminationConfigurational2018} or by introducing frustration that conversely increases the configurational entropy~\cite{chakrabartyDynamicsGlassForming2015}. In the first case, the Kauzmann temperature should increase with the disorder strength, while it should decrease in the second case. In our model, $T_0$ decreases with the increase of disorder. However, the generality of this result, and its dependence on the particular kind of disorder introduced, call for further investigations.

In conclusion, our study identifies 2D vortex lattices forming in type II superconductors as an ideal platform to investigate the fragile-to-strong glass transition. On the one hand, they offer good control over the level of intrinsic disorder present in the sample, which can be increased by reducing the film thickness or artificially engineered, e.g. by building ordered nano-pores geometries where the SC order parameter is suppressed, see \cite{Bose2023} and references therein. Most importantly, the nature of the glass-former liquid can be controlled by an external parameter, i.e. the magnetic field, enabling a systematic study of the interplay between disorder, orientational correlations, and vortex density. 
Beyond 2D thin superconducting films, layered 3D superconductors may also provide new insights into the nature of the glass transition, including the evolution of the vortex lattice fragility through the 2D to 3D crossover.

\section{Methods}

\textbf{Details of Monte Carlo numerical simulations}.
To extract the static properties of the system, we performed $10^6$ MC steps, each one consisting of $32$ Metropolis-Hastings updates of the whole lattice followed by one micro-canonical overrelaxation update of all the spins. To speed up the thermalization at lower temperatures, we performed a parallel tempering swap at each MC step. 
For each temperature and disorder strength, we discarded the transient time, which typically occurs after the first $3\times10^5$ MC steps. 
The mean value of each observable is then obtained from an average over the MC time, with a Bootstrap resampling to estimate the statistical error, combined with an average over $N_{samples}=15$ different independent disorder realizations. 

For the dynamical observable, we needed to follow a different thermalization procedure to make the MC dynamics realistic and to follow each vortex in time. To this aim, we could not use the over-relaxation nor the parallel tempering algorithm.
For each temperature and disordered configuration, we performed $2 \times 10^8$  MC steps, where a single MC step consists of a single local Metropolis-Hastings update of all the lattice spins. To help the system thermalize at low temperatures, we used a simulated Annealing procedure. For each simulation, we discarded the first ${t}_{transient}$ MC steps needed for the system to relax to the equilibrium. We fixed a threshold for ${t}_{transient}$ to $10^8$ MC steps. Finally, also in this case we additionally averaged the observables over $N_{samples}=10$ independent disorder realizations.

\begin{acknowledgement}
We would like to thank Dragana Popovi\'c for sharing the experimental data presented in Ref~\cite{maccariTransportSignaturesFragile2023} and P. Raychaudhuri for useful discussions.
I.M. acknowledges the Swiss National Science Foundation postdoctoral grant TMPFP2{\_}217204. L.B. acknowledges financial support by EU under project MORE-TEM ERC-Syn grant agreement No. 951215 and by Sapienza University of Rome under Project Ateneo 2022 RP1221816662A977 and Project Ateneo 2023 RM123188E357C540. J. L. acknowledges financial support through projects: PRIN 20207ZXT4Z QT-FLUO 
and CNR-CONICET 2023-2024 NMES. C.D.M. acknowledges financial support from the European Union --- Next Generation EU (MUR-PRIN2022 TAMeQUAD
CUP:B53D23004500006) and from ICSC---Centro Nazionale di Ricerca in High Performance Computing, Big Data, and Quantum Computing, funded by the European Union–NextGenerationEU.

\end{acknowledgement}

\begin{suppinfo}

Details on the vortex detection within a 2D spin model, derivation of the superfluid stiffness, and additional dynamical observables including the dynamical autocorrelation function and non-Gaussian parameter.
% This will usually read something like: ``Experimental procedures and
% characterization data for all new compounds. The class will
% automatically add a sentence pointing to the information on-line:

\end{suppinfo}

%\bibliography{My_Library_2024}
\clearpage
\section{\centering Supporting Information for "Fragile-to-strong glass transition in two-dimensional vortex liquids"}

\subsection{Vortex detection in a spin model with a transverse magnetic field}

The position of the vortices is determined by computing the gauge-invariant phase circulation around each unitary plaquette of the square spin grid, being

\begin{equation}
    \begin{split}
    2\pi v_i &= \left[ \theta_i-\theta_{i+\hat{x}}+F^x_i \right]^{\pi}_{-\pi}  + \left[ \theta_{i+\hat{x}}-\theta_{i+\hat{x} + \hat{y}}+F^y_{i+\hat{x}} \right]^{\pi}_{-\pi} + \\
    &+ \left[ \theta_{i+\hat{x} + \hat{y}}-\theta_{i+ \hat{y}} -F^x_{i+\hat{x} + \hat{y}} \right]^{\pi}_{-\pi} + \left[ \theta_{i + \hat{y}}-\theta_{i} -F^y_{i + \hat{y}} \right]^{\pi}_{-\pi},
    \end{split}
\end{equation}
where $\left[ \dots \right]^{\pi}_{-\pi}$ indicates that the phase difference across each bond is folded in the interval $(-\pi, \pi]$, and $v_i=\pm 1$ indicates that a vortex ($+1$) or an antivortex ($-1$) is located at the position $i$. As already mentioned, in the presence of an external magnetic field the ground state is formed by a fixed number of vortices with a given vorticity determined by the sign of the $H$ field itself. 

\subsection{Superfluid Stiffness}

In linear response theory, the expression of the  superfluid stiffness  $J_s^{\mu}= -\frac{1}{L^2} \frac{\partial^2 \ln Z(\Delta_{\mu})}{\partial \Delta_{\mu}^2} \Big|_{\Delta_{\mu}=0}$ computed along the direction $\mu$ reads

\begin{equation}
\begin{split}
J_s =  \frac{1}{L^2}& \left\langle \sum_{i} J_i^{\mu} \cos(\theta_i-\theta_{i+\hat{\mu}}+F^{\mu}_i) \right\rangle
+ \\&- \Bigg\{  \frac{1}{TL^2} \left\langle \left[\sum_i J_i^\mu \sin(\theta_i-\theta_{i+\hat{\mu}}+F^\mu_i)\right]^2 \right\rangle - \frac{1}{TL^2} \left\langle \sum_i J_i^\mu \sin(\theta_i-\theta_{i+\hat{\mu}}+F^\mu_i) \right\rangle^2 \Bigg\}.
\end{split}
\label{js}
\end{equation}

\subsection{Dynamical autocorrelation function and non-Gaussian parameter}
To better characterize the dynamical properties of the vortex lattice, we additionally computed the dynamical autocorrelation function and the non-Gaussian parameter. 

The dynamical autocorrelation function is defined as the self-part of the intermediate scattering function, it reads

\be
F_s(|\mathbf{k}_{max}|, t) =  \frac{1}{N_{\mathbf{k}}} \sum_{\mathbf{k} : |\mathbf{k}| = |\mathbf{k}^*|} \langle \exp\{i \mathbf{k} [ \mathbf{r}_j(t_0 +t) - \mathbf{r}_j(t_0)] \rangle,
\label{Self}
\ee 
where $\langle \dots \rangle$ stays, as usual, for the thermal average and the average over 10 independent disorder realizations, while $|\mathbf{k}_{max}|= 2\pi/a_\mathrm{v}$, with  $a_\mathrm{v}$ the vortex lattice spacing, is the reciprocal vector at which the structure factor shows its first peak, and $\mathbf{r}_j(t)$ is the position of the j-$th$ vortex at the MC time $t$. 

The resulting curves for different temperatures and disorder strengths are shown in Fig.~\ref{fig:Self}. $F_s(|\mathbf{k}_{max}|, t)$ provides direct information on the system relaxation to equilibrium. Liquids typically show an exponential decay with a single autocorrelation time $\tau$. Differently, glass-former supercooled liquids show a two-step function, where the second step rather than an exponential decay follows a stretched exponential function. 

At high temperatures, and for each level of disorder, the vortex lattice shows the typical liquid behaviour. However, by lowering the temperature the profile of the autocorrelation function qualitatively changes showing a two-step relaxation character whose characteristics depend on the disorder level. At low disorder, the two-step profile emerges by approaching the hexatic phase where a plateau in $t_{MC}$ appears. 
By increasing the disorder level the observed plateau shrinks and acquires an almost linear dependence on $t_\mathrm{MC}$. This effect signals a change in the glass-forming liquid nature. While at low disorder the onset of a finite orientational order in the hexatic phase constrains the dynamics leading to a pronounced plateau, the increase of disorder induces a different kind of particle caging that competes with the hexatic one eventually destroying the orientational correlations for large enough $\sigma$.

\begin{figure*}[h!]
    \includegraphics[width=\linewidth]{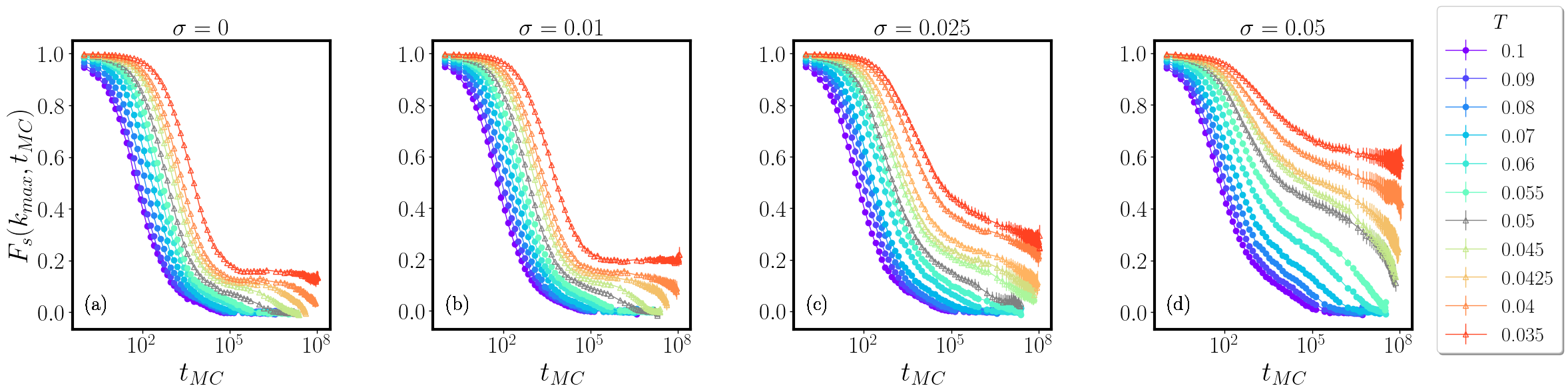}
    \caption{Dynamical autocorrelation function of the vortex lattice defined in Eq.\eqref{Self} for different temperatures and disorder levels. The value of $\sigma$ in the different panels is respectively: (a) $\sigma=0$, (b) $\sigma=0.01$, (c) $\sigma=0.025$, and (d) $\sigma=0.05$. Analogously to Fig.\ref{fig:msd_eta}, the gray lines denote the hexatic temperature $T_{hex}\simeq 0.05$.}
    \label{fig:Self}
\end{figure*}

The non-Gaussian parameter reads
\be
\alpha_2(t)= \frac{1}{2} \frac{\left \langle \Delta \mathbf{r}^4(t) \right\rangle }{\left \langle \Delta \mathbf{r}^2(t) \right\rangle^2 } - 1.
\label{alpha}
\ee
In the presence of homogeneous dynamics $\alpha_2(t)=0$, finite values of $\alpha_2(t)$ thus quantify the heterogeneity of the dynamics both in terms of strength and time extension~\cite{rahmanCorrelationsMotionAtoms1964}.
The resulting curves for different temperatures and disorder strengths are shown in Fig.\ref{fig:alpha}. While at short time scales, i.e. $t_{MC}< 10^4$ not shown in the plots, $\alpha_2(t)$ reflects the dynamic heterogeneity due to the presence of a squared numerical grid, at large time scales it provides information on the vortex-vortex interactions and the nature of the dynamics in the supercooled liquid phase. 

\begin{figure}[h!]
    \includegraphics[width=\linewidth]{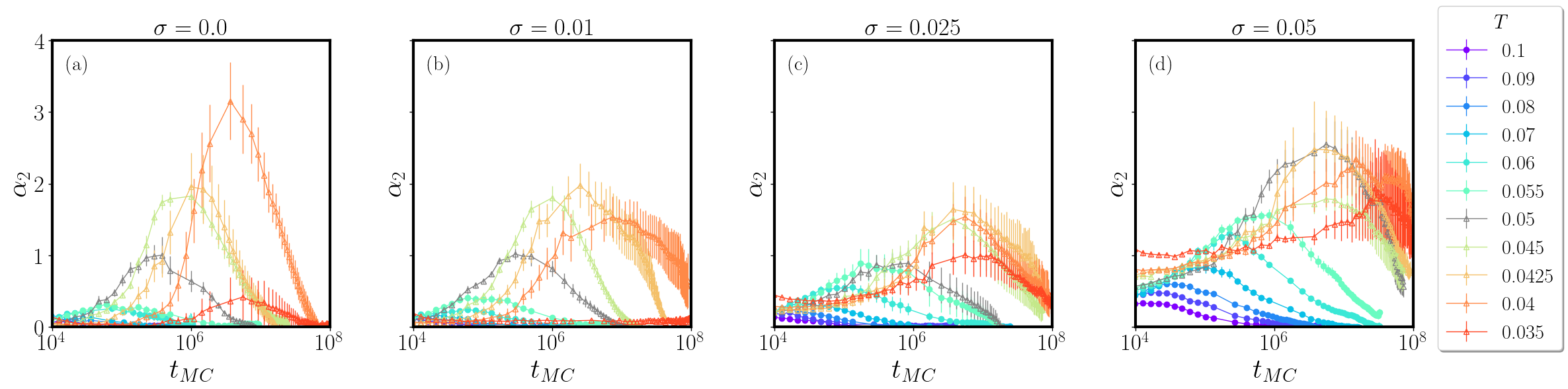}
    \caption{Non-Gaussian parameter $\alpha_2$ defined in Eq.\eqref{alpha} as a function of the MC times for different temperatures and at different disorder levels. More specifically, the value of $\sigma$ in the different panels is respectively: (a) $\sigma=0.01$, (b) $\sigma=0.025$, and (c) $\sigma=0.05$. }
    \label{fig:alpha}
\end{figure}

At low disorder, strong signatures of heterogeneous dynamics appear for temperatures $T<T_{hex}$, within the hexatic liquid phase. The onset of a dome at large time scales signals the entrance into a hexatic cage while the peak of the dome corresponds to the time at which vortices escape from it reaching the highest level of heterogeneity in their dynamics. 
By increasing the disorder we observe a reentrant behaviour reflecting a change in the origin of the heterogenous dynamics. Indeed, while up to $\sigma=0.025$ the increase of disorder acts by reducing the strength of the heterogeneity and broadening the peak of the hexatic cage, for $\sigma=0.05$ the dynamics again become strongly heterogeneous despite the vanishing of the hexatic order parameter $G_6$. 
This can be understood as a disorder-driven caging mechanism which competes with the orientational order and additionally leads to an increase of the dynamics heterogeneity at lower temperatures

\bibliography{My_Library_2024}

\providecommand{\latin}[1]{#1}
\makeatletter
\providecommand{\doi}
  {\begingroup\let\do\@makeother\dospecials
  \catcode`\{=1 \catcode`\}=2 \doi@aux}
\providecommand{\doi@aux}[1]{\endgroup\texttt{#1}}
\makeatother
\providecommand*\mcitethebibliography{\thebibliography}
\csname @ifundefined\endcsname{endmcitethebibliography}
  {\let\endmcitethebibliography\endthebibliography}{}
\begin{mcitethebibliography}{47}
\providecommand*\natexlab[1]{#1}
\providecommand*\mciteSetBstSublistMode[1]{}
\providecommand*\mciteSetBstMaxWidthForm[2]{}
\providecommand*\mciteBstWouldAddEndPuncttrue
  {\def\EndOfBibitem{\unskip.}}
\providecommand*\mciteBstWouldAddEndPunctfalse
  {\let\EndOfBibitem\relax}
\providecommand*\mciteSetBstMidEndSepPunct[3]{}
\providecommand*\mciteSetBstSublistLabelBeginEnd[3]{}
\providecommand*\EndOfBibitem{}
\mciteSetBstSublistMode{f}
\mciteSetBstMaxWidthForm{subitem}{(\alph{mcitesubitemcount})}
\mciteSetBstSublistLabelBeginEnd
  {\mcitemaxwidthsubitemform\space}
  {\relax}
  {\relax}

\bibitem[Kauzmann(1948)]{Kauzmann1948}
Kauzmann,~W. The Nature of the Glassy State and the Behavior of Liquids at Low
  Temperatures. \emph{Chemical Reviews} \textbf{1948}, \emph{43},
  219--256\relax
\mciteBstWouldAddEndPuncttrue
\mciteSetBstMidEndSepPunct{\mcitedefaultmidpunct}
{\mcitedefaultendpunct}{\mcitedefaultseppunct}\relax
\EndOfBibitem
\bibitem[Angell(1995)]{Angell1995}
Angell,~C.~A. Formation of Glasses from Liquids and Biopolymers. \emph{Science}
  \textbf{1995}, \emph{267}, 1924--1935\relax
\mciteBstWouldAddEndPuncttrue
\mciteSetBstMidEndSepPunct{\mcitedefaultmidpunct}
{\mcitedefaultendpunct}{\mcitedefaultseppunct}\relax
\EndOfBibitem
\bibitem[Vogel(1921)]{Vogel1921}
Vogel,~H. The temperature dependence law of the viscosity of fluids.
  \emph{Phys. Z.} \textbf{1921}, \emph{22}\relax
\mciteBstWouldAddEndPuncttrue
\mciteSetBstMidEndSepPunct{\mcitedefaultmidpunct}
{\mcitedefaultendpunct}{\mcitedefaultseppunct}\relax
\EndOfBibitem
\bibitem[Fulcher(1925)]{Fulcher1925}
Fulcher,~G.~S. ANALYSIS OF RECENT MEASUREMENTS OF THE VISCOSITY OF GLASSES.
  \emph{Journal of the American Ceramic Society} \textbf{1925}, \emph{8},
  339--355\relax
\mciteBstWouldAddEndPuncttrue
\mciteSetBstMidEndSepPunct{\mcitedefaultmidpunct}
{\mcitedefaultendpunct}{\mcitedefaultseppunct}\relax
\EndOfBibitem
\bibitem[Tammann and Hesse(1926)Tammann, and Hesse]{Tamman1926}
Tammann,~G.; Hesse,~W. Die Abhängigkeit der Viscosität von der Temperatur bie
  unterkühlten Flüssigkeiten. \emph{Zeitschrift für anorganische und
  allgemeine Chemie} \textbf{1926}, \emph{156}, 245--257\relax
\mciteBstWouldAddEndPuncttrue
\mciteSetBstMidEndSepPunct{\mcitedefaultmidpunct}
{\mcitedefaultendpunct}{\mcitedefaultseppunct}\relax
\EndOfBibitem
\bibitem[Cavagna(2009)]{cavagnaSupercooledLiquidsPedestrians2009}
Cavagna,~A. Supercooled Liquids for Pedestrians. \emph{Physics Reports}
  \textbf{2009}, \emph{476}, 51--124\relax
\mciteBstWouldAddEndPuncttrue
\mciteSetBstMidEndSepPunct{\mcitedefaultmidpunct}
{\mcitedefaultendpunct}{\mcitedefaultseppunct}\relax
\EndOfBibitem
\bibitem[Ruocco \latin{et~al.}(2004)Ruocco, Sciortino, Zamponi, De~Michele, and
  Scopigno]{Ruocco2004}
Ruocco,~G.; Sciortino,~F.; Zamponi,~F.; De~Michele,~C.; Scopigno,~T. Landscapes
  and fragilities. \emph{Journal of Chemical Physics} \textbf{2004},
  \emph{120}, 10666--10680, cited By 77\relax
\mciteBstWouldAddEndPuncttrue
\mciteSetBstMidEndSepPunct{\mcitedefaultmidpunct}
{\mcitedefaultendpunct}{\mcitedefaultseppunct}\relax
\EndOfBibitem
\bibitem[Raymond P.~Seekell \latin{et~al.}(2015)Raymond P.~Seekell,
  Sarangapani, Zhang, and
  Zhu]{raymondp.seekellRelationshipParticleElasticity2015a}
Raymond P.~Seekell,~I. I.~I.; Sarangapani,~P.~S.; Zhang,~Z.; Zhu,~Y.
  Relationship between Particle Elasticity, Glass Fragility, and Structural
  Relaxation in Dense Microgel Suspensions. \emph{Soft Matter} \textbf{2015},
  \emph{11}, 5485--5491\relax
\mciteBstWouldAddEndPuncttrue
\mciteSetBstMidEndSepPunct{\mcitedefaultmidpunct}
{\mcitedefaultendpunct}{\mcitedefaultseppunct}\relax
\EndOfBibitem
\bibitem[Nigro \latin{et~al.}(2017)Nigro, Angelini, Bertoldo, Bruni, Ricci, and
  Ruzicka]{nigroDynamicalBehaviorMicrogels2017}
Nigro,~V.; Angelini,~R.; Bertoldo,~M.; Bruni,~F.; Ricci,~M.~A.; Ruzicka,~B.
  Dynamical Behavior of Microgels of Interpenetrated Polymer Networks.
  \emph{Soft Matter} \textbf{2017}, \emph{13}, 5185--5193\relax
\mciteBstWouldAddEndPuncttrue
\mciteSetBstMidEndSepPunct{\mcitedefaultmidpunct}
{\mcitedefaultendpunct}{\mcitedefaultseppunct}\relax
\EndOfBibitem
\bibitem[Nigro \latin{et~al.}(2020)Nigro, Ruzicka, Ruta, Zontone, Bertoldo,
  Buratti, and Angelini]{nigroRelaxationDynamicsSoftness2020}
Nigro,~V.; Ruzicka,~B.; Ruta,~B.; Zontone,~F.; Bertoldo,~M.; Buratti,~E.;
  Angelini,~R. Relaxation {{Dynamics}}, {{Softness}}, and {{Fragility}} of
  {{Microgels}} with {{Interpenetrated Polymer Networks}}.
  \emph{Macromolecules} \textbf{2020}, \emph{53}, 1596--1603\relax
\mciteBstWouldAddEndPuncttrue
\mciteSetBstMidEndSepPunct{\mcitedefaultmidpunct}
{\mcitedefaultendpunct}{\mcitedefaultseppunct}\relax
\EndOfBibitem
\bibitem[Mattsson \latin{et~al.}(2009)Mattsson, Wyss, {Fernandez-Nieves},
  Miyazaki, Hu, Reichman, and Weitz]{mattssonSoftColloidsMake2009}
Mattsson,~J.; Wyss,~H.~M.; {Fernandez-Nieves},~A.; Miyazaki,~K.; Hu,~Z.;
  Reichman,~D.~R.; Weitz,~D.~A. Soft Colloids Make Strong Glasses.
  \emph{Nature} \textbf{2009}, \emph{462}, 83--86\relax
\mciteBstWouldAddEndPuncttrue
\mciteSetBstMidEndSepPunct{\mcitedefaultmidpunct}
{\mcitedefaultendpunct}{\mcitedefaultseppunct}\relax
\EndOfBibitem
\bibitem[van~der Scheer \latin{et~al.}(2017)van~der Scheer, van~de Laar,
  van~der Gucht, Vlassopoulos, and Sprakel]{vanderScheerFragility2017}
van~der Scheer,~P.; van~de Laar,~T.; van~der Gucht,~J.; Vlassopoulos,~D.;
  Sprakel,~J. Fragility and Strength in Nanoparticle Glasses. \emph{ACS Nano}
  \textbf{2017}, \emph{11}, 6755--6763, PMID: 28658568\relax
\mciteBstWouldAddEndPuncttrue
\mciteSetBstMidEndSepPunct{\mcitedefaultmidpunct}
{\mcitedefaultendpunct}{\mcitedefaultseppunct}\relax
\EndOfBibitem
\bibitem[Yu \latin{et~al.}(2022)Yu, Morgan, Ediger, and Wang]{Yu2022}
Yu,~Z.; Morgan,~D.; Ediger,~M.~D.; Wang,~B. Understanding the Fragile-to-Strong
  Transition in Silica from Microscopic Dynamics. \emph{Phys. Rev. Lett.}
  \textbf{2022}, \emph{129}, 018003\relax
\mciteBstWouldAddEndPuncttrue
\mciteSetBstMidEndSepPunct{\mcitedefaultmidpunct}
{\mcitedefaultendpunct}{\mcitedefaultseppunct}\relax
\EndOfBibitem
\bibitem[Gnan and Zaccarelli(2019)Gnan, and
  Zaccarelli]{gnanMicroscopicRoleDeformation2019}
Gnan,~N.; Zaccarelli,~E. The Microscopic Role of Deformation in the Dynamics of
  Soft Colloids. \emph{Nature Physics} \textbf{2019}, \emph{15}, 683--688\relax
\mciteBstWouldAddEndPuncttrue
\mciteSetBstMidEndSepPunct{\mcitedefaultmidpunct}
{\mcitedefaultendpunct}{\mcitedefaultseppunct}\relax
\EndOfBibitem
\bibitem[Berezinsky(1972)]{berezinskyDestructionLongrangeOrder1972}
Berezinsky,~V.~L. Destruction of {{Long-range Order}} in {{One-dimensional}}
  and {{Two-dimensional Systems Possessing}} a {{Continuous Symmetry Group}}.
  {{II}}. {{Quantum Systems}}. \emph{Sov. Phys. JETP} \textbf{1972}, \emph{34},
  610\relax
\mciteBstWouldAddEndPuncttrue
\mciteSetBstMidEndSepPunct{\mcitedefaultmidpunct}
{\mcitedefaultendpunct}{\mcitedefaultseppunct}\relax
\EndOfBibitem
\bibitem[Kosterlitz and Thouless(1973)Kosterlitz, and
  Thouless]{kosterlitzOrderingMetastabilityPhase1973}
Kosterlitz,~J.~M.; Thouless,~D.~J. Ordering, Metastability and Phase
  Transitions in Two-Dimensional Systems. \emph{Journal of Physics C: Solid
  State Physics} \textbf{1973}, \emph{6}, 1181--1203\relax
\mciteBstWouldAddEndPuncttrue
\mciteSetBstMidEndSepPunct{\mcitedefaultmidpunct}
{\mcitedefaultendpunct}{\mcitedefaultseppunct}\relax
\EndOfBibitem
\bibitem[Kosterlitz(1974)]{kosterlitzCriticalPropertiesTwodimensional1974}
Kosterlitz,~J.~M. The Critical Properties of the Two-Dimensional Xy Model.
  \emph{Journal of Physics C: Solid State Physics} \textbf{1974}, \emph{7},
  1046\relax
\mciteBstWouldAddEndPuncttrue
\mciteSetBstMidEndSepPunct{\mcitedefaultmidpunct}
{\mcitedefaultendpunct}{\mcitedefaultseppunct}\relax
\EndOfBibitem
\bibitem[Halperin and Nelson(1978)Halperin, and
  Nelson]{halperinTheoryTwoDimensionalMelting1978}
Halperin,~B.~I.; Nelson,~D.~R. Theory of {{Two-Dimensional Melting}}.
  \emph{Physical Review Letters} \textbf{1978}, \emph{41}, 121--124\relax
\mciteBstWouldAddEndPuncttrue
\mciteSetBstMidEndSepPunct{\mcitedefaultmidpunct}
{\mcitedefaultendpunct}{\mcitedefaultseppunct}\relax
\EndOfBibitem
\bibitem[Nelson and Halperin(1979)Nelson, and
  Halperin]{nelsonDislocationmediatedMeltingTwo1979}
Nelson,~D.~R.; Halperin,~B.~I. Dislocation-Mediated Melting in Two Dimensions.
  \emph{Physical Review B} \textbf{1979}, \emph{19}, 2457--2484\relax
\mciteBstWouldAddEndPuncttrue
\mciteSetBstMidEndSepPunct{\mcitedefaultmidpunct}
{\mcitedefaultendpunct}{\mcitedefaultseppunct}\relax
\EndOfBibitem
\bibitem[Young(1979)]{youngMeltingVectorCoulomb1979}
Young,~A.~P. Melting and the Vector {{Coulomb}} Gas in Two Dimensions.
  \emph{Physical Review B} \textbf{1979}, \emph{19}, 1855--1866\relax
\mciteBstWouldAddEndPuncttrue
\mciteSetBstMidEndSepPunct{\mcitedefaultmidpunct}
{\mcitedefaultendpunct}{\mcitedefaultseppunct}\relax
\EndOfBibitem
\bibitem[Zangi and Rice(2004)Zangi, and Rice]{zangiCooperativeDynamicsTwo2004}
Zangi,~R.; Rice,~S.~A. Cooperative {{Dynamics}} in {{Two Dimensions}}.
  \emph{Physical Review Letters} \textbf{2004}, \emph{92}, 035502\relax
\mciteBstWouldAddEndPuncttrue
\mciteSetBstMidEndSepPunct{\mcitedefaultmidpunct}
{\mcitedefaultendpunct}{\mcitedefaultseppunct}\relax
\EndOfBibitem
\bibitem[Shiba \latin{et~al.}(2009)Shiba, Onuki, and
  Araki]{shibaStructuralDynamicalHeterogeneities2009}
Shiba,~H.; Onuki,~A.; Araki,~T. Structural and Dynamical Heterogeneities in
  Two-Dimensional Melting. \emph{EPL (Europhysics Letters)} \textbf{2009},
  \emph{86}, 66004\relax
\mciteBstWouldAddEndPuncttrue
\mciteSetBstMidEndSepPunct{\mcitedefaultmidpunct}
{\mcitedefaultendpunct}{\mcitedefaultseppunct}\relax
\EndOfBibitem
\bibitem[Kawasaki and Tanaka(2011)Kawasaki, and
  Tanaka]{kawasakiStructuralSignatureSlow2011}
Kawasaki,~T.; Tanaka,~H. Structural Signature of Slow Dynamics and Dynamic
  Heterogeneity in Two-Dimensional Colloidal Liquids: Glassy Structural Order.
  \emph{Journal of Physics: Condensed Matter} \textbf{2011}, \emph{23},
  194121\relax
\mciteBstWouldAddEndPuncttrue
\mciteSetBstMidEndSepPunct{\mcitedefaultmidpunct}
{\mcitedefaultendpunct}{\mcitedefaultseppunct}\relax
\EndOfBibitem
\bibitem[Bernard and Krauth(2011)Bernard, and Krauth]{Krauth2011}
Bernard,~E.~P.; Krauth,~W. Two-Step Melting in Two Dimensions: First-Order
  Liquid-Hexatic Transition. \emph{Phys. Rev. Lett.} \textbf{2011}, \emph{107},
  155704\relax
\mciteBstWouldAddEndPuncttrue
\mciteSetBstMidEndSepPunct{\mcitedefaultmidpunct}
{\mcitedefaultendpunct}{\mcitedefaultseppunct}\relax
\EndOfBibitem
\bibitem[Kapfer and Krauth(2015)Kapfer, and Krauth]{Krauth2015}
Kapfer,~S.~C.; Krauth,~W. Two-Dimensional Melting: From Liquid-Hexatic
  Coexistence to Continuous Transitions. \emph{Phys. Rev. Lett.} \textbf{2015},
  \emph{114}, 035702\relax
\mciteBstWouldAddEndPuncttrue
\mciteSetBstMidEndSepPunct{\mcitedefaultmidpunct}
{\mcitedefaultendpunct}{\mcitedefaultseppunct}\relax
\EndOfBibitem
\bibitem[van~der Meer \latin{et~al.}(2015)van~der Meer, Qi, Sprakel, Filion,
  and Dijkstra]{meerDynamicalHeterogeneitiesDefects2015}
van~der Meer,~B.; Qi,~W.; Sprakel,~J.; Filion,~L.; Dijkstra,~M. Dynamical
  Heterogeneities and Defects in Two-Dimensional Soft Colloidal Crystals.
  \emph{Soft Matter} \textbf{2015}, \emph{11}, 9385--9392\relax
\mciteBstWouldAddEndPuncttrue
\mciteSetBstMidEndSepPunct{\mcitedefaultmidpunct}
{\mcitedefaultendpunct}{\mcitedefaultseppunct}\relax
\EndOfBibitem
\bibitem[Tah \latin{et~al.}(2018)Tah, Sengupta, Sastry, Dasgupta, and
  Karmakar]{tahGlassTransitionSupercooled2018}
Tah,~I.; Sengupta,~S.; Sastry,~S.; Dasgupta,~C.; Karmakar,~S. Glass
  {{Transition}} in {{Supercooled Liquids}} with {{Medium-Range Crystalline
  Order}}. \emph{Physical Review Letters} \textbf{2018}, \emph{121},
  085703\relax
\mciteBstWouldAddEndPuncttrue
\mciteSetBstMidEndSepPunct{\mcitedefaultmidpunct}
{\mcitedefaultendpunct}{\mcitedefaultseppunct}\relax
\EndOfBibitem
\bibitem[Giamarchi and Le~Doussal(1995)Giamarchi, and
  Le~Doussal]{giamarchiElasticTheoryFlux1995}
Giamarchi,~T.; Le~Doussal,~P. Elastic Theory of Flux Lattices in the Presence
  of Weak Disorder. \emph{Physical Review B} \textbf{1995}, \emph{52},
  1242--1270\relax
\mciteBstWouldAddEndPuncttrue
\mciteSetBstMidEndSepPunct{\mcitedefaultmidpunct}
{\mcitedefaultendpunct}{\mcitedefaultseppunct}\relax
\EndOfBibitem
\bibitem[Maccari \latin{et~al.}(2023)Maccari, Pokharel, Terzic, Dutta,
  Jesudasan, Raychaudhuri, Lorenzana, De~Michele, Castellani, Benfatto, and
  Popovi{\'c}]{maccariTransportSignaturesFragile2023}
Maccari,~I.; Pokharel,~B.~K.; Terzic,~J.; Dutta,~S.; Jesudasan,~J.;
  Raychaudhuri,~P.; Lorenzana,~J.; De~Michele,~C.; Castellani,~C.;
  Benfatto,~L.; Popovi{\'c},~D. Transport Signatures of Fragile Glass Dynamics
  in the Melting of the Two-Dimensional Vortex Lattice. \emph{Physical Review
  B} \textbf{2023}, \emph{107}, 014509\relax
\mciteBstWouldAddEndPuncttrue
\mciteSetBstMidEndSepPunct{\mcitedefaultmidpunct}
{\mcitedefaultendpunct}{\mcitedefaultseppunct}\relax
\EndOfBibitem
\bibitem[Roy \latin{et~al.}(2019)Roy, Dutta, Roy~Choudhury, Basistha, Maccari,
  Mandal, Jesudasan, Bagwe, Castellani, Benfatto, and
  Raychaudhuri]{royMeltingVortexLattice2019}
Roy,~I.; Dutta,~S.; Roy~Choudhury,~A.~N.; Basistha,~S.; Maccari,~I.;
  Mandal,~S.; Jesudasan,~J.; Bagwe,~V.; Castellani,~C.; Benfatto,~L.;
  Raychaudhuri,~P. Melting of the {{Vortex Lattice}} through {{Intermediate
  Hexatic Fluid}} in an a - {{MoGe Thin Film}}. \emph{Physical Review Letters}
  \textbf{2019}, \emph{122}, 047001\relax
\mciteBstWouldAddEndPuncttrue
\mciteSetBstMidEndSepPunct{\mcitedefaultmidpunct}
{\mcitedefaultendpunct}{\mcitedefaultseppunct}\relax
\EndOfBibitem
\bibitem[Dutta \latin{et~al.}(2019)Dutta, Roy, Mandal, Jesudasan, Bagwe, and
  Raychaudhuri]{duttaExtremeSensitivityVortex2019}
Dutta,~S.; Roy,~I.; Mandal,~S.; Jesudasan,~J.; Bagwe,~V.; Raychaudhuri,~P.
  Extreme Sensitivity of the Vortex State in \$a\$-{{MoGe}} Films to
  Radio-Frequency Electromagnetic Perturbation. \emph{Physical Review B}
  \textbf{2019}, \emph{100}, 214518\relax
\mciteBstWouldAddEndPuncttrue
\mciteSetBstMidEndSepPunct{\mcitedefaultmidpunct}
{\mcitedefaultendpunct}{\mcitedefaultseppunct}\relax
\EndOfBibitem
\bibitem[Angell(1991)]{angellRelaxationLiquidsPolymers1991}
Angell,~C.~A. Relaxation in Liquids, Polymers and Plastic Crystals ---
  Strong/Fragile Patterns and Problems. \emph{Journal of Non-Crystalline
  Solids} \textbf{1991}, \emph{131--133}, 13--31\relax
\mciteBstWouldAddEndPuncttrue
\mciteSetBstMidEndSepPunct{\mcitedefaultmidpunct}
{\mcitedefaultendpunct}{\mcitedefaultseppunct}\relax
\EndOfBibitem
\bibitem[Plazek and Ngai(1991)Plazek, and
  Ngai]{plazekCorrelationPolymerSegmental1991}
Plazek,~D.~J.; Ngai,~K.~L. Correlation of Polymer Segmental Chain Dynamics with
  Temperature-Dependent Time-Scale Shifts. \emph{Macromolecules} \textbf{1991},
  \emph{24}, 1222--1224\relax
\mciteBstWouldAddEndPuncttrue
\mciteSetBstMidEndSepPunct{\mcitedefaultmidpunct}
{\mcitedefaultendpunct}{\mcitedefaultseppunct}\relax
\EndOfBibitem
\bibitem[B{\"o}hmer and Angell(1992)B{\"o}hmer, and
  Angell]{bohmerCorrelationsNonexponentialityState1992}
B{\"o}hmer,~R.; Angell,~C.~A. Correlations of the Nonexponentiality and State
  Dependence of Mechanical Relaxations with Bond Connectivity in {{Ge-As-Se}}
  Supercooled Liquids. \emph{Physical Review B} \textbf{1992}, \emph{45},
  10091--10094\relax
\mciteBstWouldAddEndPuncttrue
\mciteSetBstMidEndSepPunct{\mcitedefaultmidpunct}
{\mcitedefaultendpunct}{\mcitedefaultseppunct}\relax
\EndOfBibitem
\bibitem[{Saika-Voivod} \latin{et~al.}(2001){Saika-Voivod}, Poole, and
  Sciortino]{saika-voivodFragiletostrongTransitionPolyamorphism2001}
{Saika-Voivod},~I.; Poole,~P.~H.; Sciortino,~F. Fragile-to-Strong Transition
  and Polyamorphism in the Energy Landscape of Liquid Silica. \emph{Nature}
  \textbf{2001}, \emph{412}, 514--517\relax
\mciteBstWouldAddEndPuncttrue
\mciteSetBstMidEndSepPunct{\mcitedefaultmidpunct}
{\mcitedefaultendpunct}{\mcitedefaultseppunct}\relax
\EndOfBibitem
\bibitem[Maccari \latin{et~al.}(2021)Maccari, Benfatto, and
  Castellani]{maccariUniformlyFrustratedXY2021}
Maccari,~I.; Benfatto,~L.; Castellani,~C. Uniformly {{Frustrated XY Model}}:
  {{Strengthening}} of the {{Vortex Lattice}} by {{Intrinsic Disorder}}.
  \emph{Condensed Matter} \textbf{2021}, \emph{6}, 42\relax
\mciteBstWouldAddEndPuncttrue
\mciteSetBstMidEndSepPunct{\mcitedefaultmidpunct}
{\mcitedefaultendpunct}{\mcitedefaultseppunct}\relax
\EndOfBibitem
\bibitem[Deutschl{\"a}nder \latin{et~al.}(2013)Deutschl{\"a}nder, Horn,
  L{\"o}wen, Maret, and Keim]{deutschlanderTwoDimensionalMeltingQuenched2013}
Deutschl{\"a}nder,~S.; Horn,~T.; L{\"o}wen,~H.; Maret,~G.; Keim,~P.
  Two-{{Dimensional Melting}} under {{Quenched Disorder}}. \emph{Physical
  Review Letters} \textbf{2013}, \emph{111}, 098301\relax
\mciteBstWouldAddEndPuncttrue
\mciteSetBstMidEndSepPunct{\mcitedefaultmidpunct}
{\mcitedefaultendpunct}{\mcitedefaultseppunct}\relax
\EndOfBibitem
\bibitem[E.~A.~Gaiduk and Ryzhov(2019)E.~A.~Gaiduk, and Ryzhov]{Gaiduk2019}
E.~A.~Gaiduk,~E. N.~T.,~Yu.D.~Fomin; Ryzhov,~V.~N. The influence of random
  pinning on the melting scenario of two-dimensional soft-disk systems.
  \emph{Molecular Physics} \textbf{2019}, \emph{117}, 2910--2919\relax
\mciteBstWouldAddEndPuncttrue
\mciteSetBstMidEndSepPunct{\mcitedefaultmidpunct}
{\mcitedefaultendpunct}{\mcitedefaultseppunct}\relax
\EndOfBibitem
\bibitem[Shankaraiah \latin{et~al.}(2019)Shankaraiah, Sengupta, and
  Menon]{Shankaraiah2019}
Shankaraiah,~N.; Sengupta,~S.; Menon,~G.~I. {Orientational correlations in
  fluids with quenched disorder}. \emph{The Journal of Chemical Physics}
  \textbf{2019}, \emph{151}, 124501\relax
\mciteBstWouldAddEndPuncttrue
\mciteSetBstMidEndSepPunct{\mcitedefaultmidpunct}
{\mcitedefaultendpunct}{\mcitedefaultseppunct}\relax
\EndOfBibitem
\bibitem[Berthier and Kob(2007)Berthier, and Kob]{Kob2007}
Berthier,~L.; Kob,~W. The Monte Carlo dynamics of a binary Lennard-Jones
  glass-forming mixture. \emph{Journal of Physics: Condensed Matter}
  \textbf{2007}, \emph{19}, 205130\relax
\mciteBstWouldAddEndPuncttrue
\mciteSetBstMidEndSepPunct{\mcitedefaultmidpunct}
{\mcitedefaultendpunct}{\mcitedefaultseppunct}\relax
\EndOfBibitem
\bibitem[Sentjabrskaja \latin{et~al.}(2016)Sentjabrskaja, Zaccarelli, Michele,
  Sciortino, Tartaglia, Voigtmann, Egelhaaf, and
  Laurati]{Sentjabrskaja2016AnomalousDO}
Sentjabrskaja,~T.; Zaccarelli,~E.; Michele,~C.~D.; Sciortino,~F.;
  Tartaglia,~P.; Voigtmann,~T.; Egelhaaf,~S.~U.; Laurati,~M. Anomalous dynamics
  of intruders in a crowded environment of mobile obstacles. \emph{Nature
  Communications} \textbf{2016}, \emph{7}\relax
\mciteBstWouldAddEndPuncttrue
\mciteSetBstMidEndSepPunct{\mcitedefaultmidpunct}
{\mcitedefaultendpunct}{\mcitedefaultseppunct}\relax
\EndOfBibitem
\bibitem[Cammarota and Biroli(2012)Cammarota, and
  Biroli]{cammarotaIdealGlassTransitions2012}
Cammarota,~C.; Biroli,~G. Ideal Glass Transitions by Random Pinning.
  \emph{Proceedings of the National Academy of Sciences} \textbf{2012},
  \emph{109}, 8850--8855\relax
\mciteBstWouldAddEndPuncttrue
\mciteSetBstMidEndSepPunct{\mcitedefaultmidpunct}
{\mcitedefaultendpunct}{\mcitedefaultseppunct}\relax
\EndOfBibitem
\bibitem[Williams \latin{et~al.}(2018)Williams, Turci, Hallett, Crowther,
  Cammarota, Biroli, and
  Royall]{williamsExperimentalDeterminationConfigurational2018}
Williams,~I.; Turci,~F.; Hallett,~J.~E.; Crowther,~P.; Cammarota,~C.;
  Biroli,~G.; Royall,~C.~P. Experimental Determination of Configurational
  Entropy in a Two-Dimensional Liquid under Random Pinning. \emph{Journal of
  Physics: Condensed Matter} \textbf{2018}, \emph{30}, 094003\relax
\mciteBstWouldAddEndPuncttrue
\mciteSetBstMidEndSepPunct{\mcitedefaultmidpunct}
{\mcitedefaultendpunct}{\mcitedefaultseppunct}\relax
\EndOfBibitem
\bibitem[Chakrabarty \latin{et~al.}(2015)Chakrabarty, Karmakar, and
  Dasgupta]{chakrabartyDynamicsGlassForming2015}
Chakrabarty,~S.; Karmakar,~S.; Dasgupta,~C. Dynamics of {{Glass Forming
  Liquids}} with {{Randomly Pinned Particles}}. \emph{Scientific Reports}
  \textbf{2015}, \emph{5}, 12577\relax
\mciteBstWouldAddEndPuncttrue
\mciteSetBstMidEndSepPunct{\mcitedefaultmidpunct}
{\mcitedefaultendpunct}{\mcitedefaultseppunct}\relax
\EndOfBibitem
\bibitem[Bose(2023)]{Bose2023}
Bose,~S. A review of superconductivity in nanostructures - from nanogranular
  films to anti-dot arrays. \emph{Superconductor Science and Technology}
  \textbf{2023}, \emph{36}, 063003\relax
\mciteBstWouldAddEndPuncttrue
\mciteSetBstMidEndSepPunct{\mcitedefaultmidpunct}
{\mcitedefaultendpunct}{\mcitedefaultseppunct}\relax
\EndOfBibitem
\bibitem[Rahman(1964)]{rahmanCorrelationsMotionAtoms1964}
Rahman,~A. Correlations in the {{Motion}} of {{Atoms}} in {{Liquid Argon}}.
  \emph{Physical Review} \textbf{1964}, \emph{136}, A405--A411\relax
\mciteBstWouldAddEndPuncttrue
\mciteSetBstMidEndSepPunct{\mcitedefaultmidpunct}
{\mcitedefaultendpunct}{\mcitedefaultseppunct}\relax
\EndOfBibitem
\end{mcitethebibliography}
\end{document}